 \documentclass[manuscript]{aastex61}

\begin{document}

\title{Gyrosynchrotron emission generated by nonthermal electrons with energy spectra of a broken power law}


\author{Zhao Wu}
\affil{Shandong Provincial Key Laboratory of Optical Astronomy and Solar-Terrestrial Environment,
and Institute of Space Sciences, Shandong University, Weihai, Shandong 264209, China; yaochen@sdu.edu.cn}

\author{Yao Chen}
\affil{Shandong Provincial Key Laboratory of Optical Astronomy and Solar-Terrestrial Environment,
and Institute of Space Sciences, Shandong University, Weihai, Shandong 264209, China; yaochen@sdu.edu.cn}

\author{Hao Ning}
\affiliation{Shandong Provincial Key Laboratory of Optical Astronomy and Solar-Terrestrial Environment,
and Institute of Space Sciences, Shandong University, Weihai, Shandong 264209, China; yaochen@sdu.edu.cn}

\author{Xiangliang Kong}
\affiliation{Shandong Provincial Key Laboratory of Optical Astronomy and Solar-Terrestrial Environment,
and Institute of Space Sciences, Shandong University, Weihai, Shandong 264209, China; yaochen@sdu.edu.cn}

\author{Jeongwoo Lee}
\affiliation{Shandong Provincial Key Laboratory of Optical Astronomy and Solar-Terrestrial Environment,
and Institute of Space Sciences, Shandong University, Weihai, Shandong 264209, China; yaochen@sdu.edu.cn}

\begin{abstract}
Latest observational reports of solar flares reveal some uncommon features of microwave spectra, such as unusually hard (or even positive) spectra, and/or a super-high peak frequency. For a better understanding of these features, we conduct a parameter study to investigate the effect of broken-power-law spectra of energetic electrons on microwave emission on the basis of gyrosynchrotron mechanism. The electron broken-power-law energy distribution is characterized by three parameters, the break energy ($E_B$), the power-law indices below ($\delta_1$) and above ($\delta_2$) the break energy. We find that with the addition of the $\delta_2$ component of the electron spectra, the total flux density can increase by several times in the optically-thick regime, and by orders of magnitude in the optically-thin regime; the peak frequency ($\nu_p$) also increases and can reach up to tens of GHz; and the degree of polarization ($r_c$) decreases in general. We also find that (1) the variation of the flux density is much larger in the optically-thin regime, and the microwave spectra around the peak frequency manifest various profiles with the softening or soft-hard pattern; (2) the parameters $\delta_1$ and $E_B$ affect the microwave spectral index ($\alpha$) and the degree of polarization ($r_c$) mainly in the optically-thick regime, while $\delta_2$ mainly affects the optically-thin regime. The results are helpful in understanding the lately-reported microwave bursts with unusual spectral features and point out the demands for a more-complete spectral coverage of microwave bursts, especially, in the high-frequency regime, say, $>10-20$ GHz.

\end{abstract}


\section{Introduction}
Microwave bursts are mainly gyrosynchrotron emission excited by mildly relativistic electrons spiraling around the magnetic fields. These electrons, accelerated during solar flares, when precipitating into the dense chromosphere, can also result in Hard X-ray (HXR) emission that is highly-correlated with the microwave emission \citep{Melrose1976,White2011,Chen2017}.

The microwave spectrum usually presents a positive slope at low frequencies (called as the optically-thick regime) and a negative slope at high frequencies (the optically-thin regime). The turn-over frequency, which separates the optically thick and thin regimes, usually appears in a range of 5-15~GHz. The most-commonly used formulae for gyrosynchrotron emission were derived by \citet{Ramaty1969}, considering the Razin suppression \citep{Razin1960,Razin1960b}, self-absorption \citep{Twiss1954} as well as maser emission \citep{Heyvaerts1968}. Using these formulae one can calculate the emissivity and the absorption coefficient at a given frequency for specified electron distributions under certain coronal conditions. The microwave flux density can then be obtained using the radiative transfer equation.

With the above formulae, observations of microwave bursts can be used to diagnose the coronal magnetic field and properties of the emitting electrons. For this purpose, the original Ramaty formulae, which are somewhat complex and time-consuming in computation, have been processed with different simplifications.

\citet{Dulk1982} and \citet{Dulk1985} presented empirical expressions for electrons with isotropic pitch angles and power-law energy distributions. With these expressions one can easily relate major microwave parameters observed (such as the peak frequency, the degree of polarization, the spectral index, and the brightness temperature) to coronal parameters (such as the magnetic field strength, the viewing angle (i.e., the angle between the wave vector and the magnetic field), the column density of energetic electrons and their power-law index). However, these simplified expressions are valid for a limited range of parameters (electron spectral index: $2\leq\delta\leq7$, viewing angle: $20^\circ\leq\theta\leq80^\circ$, harmonic numbers: $10\leq\nu/\nu_B\leq100$). A series of studies were put forward to estimate the total \citep{Zhou1994} and the vector \citep{Huang2002,Huang2006} magnetic field in microwave sources.

Another simplification has been developed by \citet{Petrosian1981} and \citet{Klein1987} (called as the PK approximation). They simplified the exact gyrosynchrotron formulae mathematically, for instance, replacing the summation over harmonics by continuous integration and using approximated forms of the Bessel functions. \citet{Fleishman2010} extended it for anisotropic electron distributions and further increased the efficiency of the computation. Along this line of approach, \citet{Gary2013} proposed a procedure to diagnose the 3D coronal magnetic field and other parameters, starting from some prescribed values of these parameters, and adjusting them by minimizing the difference between the calculated and the observed 2D maps of microwave emission so as to get the optimal diagnostic results. Similar studies were carried out by other authors \citep{Nindos2000,Kundu04b,Nita2015,Casini2017,Kuroda2018}.

Unlike standard microwave spectra as introduced above, some uncommon features have been reported. For example, in the 1984 May 21 event the emission was stronger at 90~GHz than at 30~GHz, indicating a peak frequency over 30~GHz as well as an overall positive spectrum between 30 and 90~GHz \citep{Klein1987}. In addition, events with flat or extremely-hard spectra have been reported \citep{Hachenberg1961,Raulin1999,Trottet2011,Song2016}. This indicates that the emission process may be more complicated with factors not fully considered by the above simplifications. Thus, further theoretical calculation of microwave emission is necessary. This study moves one step forward by considering energetic electrons with a broken-power-law distribution.

Existence of such broken-power-law energy distribution of energetic electrons has been demonstrated by the HXR observation showing a broken photon spectrum with a harder spectral component around 300 keV or higher \citep{Suri1972,Yoshimori1985,Shih2009,Kawate2012,Kong2013}. The straightforward interpretation of this observation is that the source electrons intrinsically have a hardening energy distribution \citep{Suri1972,Yoshimori1985,Dennis1988}. \citet{Moses1989} showed that in some flares the associated energetic electrons detected in-situ presented spectral hardening at high energies. \citet{Asai2013} found that the electron spectral index deduced from the microwave emission was harder than that from the HXRs of the same source \citep[also by][]{Silva2007}, and deduced that the source electrons may be intrinsically hard at high energies. These studies provide additional support to the above interpretation of the high-energy hardening of the associated energetic electrons.

On the other hand, theoretical investigations on electron acceleration during solar flares have managed to reproduce the spectral hardening of energetic electrons by considering the effect of termination shock \citep{Li2013,Kong2013}. The authors pointed out that higher-energy electrons ($>$300~keV) were accelerated more efficiently due to their resonance with MHD turbulence in the inertial range at the termination shock, while lower-energy electrons can only resonant with MHD turbulence in the dissipation range.

On the basis of these investigations, we assume that the energetic electrons are characterized by spectral hardening at high energies, and conduct a detailed parameter study on the effect of such electron spectra on the gyrosynchrotron microwave emission.
Next section presents parameters used in our calculations. Section 3 shows the major results. A summary and discussion are given in the last section.

\section{Parameters used to calculate the microwave emission}
\label{sec:2}
The microwave flux density and spectrum depend on three sets of parameters, including~(1) the ambient magnetic field strength and plasma electron density, (2) distribution function of energetic electrons (given by a broken-power-law spectrum of energy in the present study), and (3) the source size and the viewing angle. {The Solar SoftWare (SSW) provides a code for calculating gyrosynchrotron radiation from electrons of an isotropic pitch-angle and a single power-law distribution \citep['ramaty\_gysy\_core.pro',][]{Ramaty1994}. We modify this code to accomodate electrons with broken-power-law energy distributions.} Ranges of parameters used in our calculation are as follows.

The ambient magnetic field and plasma density are fixed to be 200~G (gyro-frequency, $\nu_B\approx560~$MHz) and 10$^{10}$~cm$^{-3}$ (plasma frequency, $\nu_0\approx898~$MHz), respectively. The Razin frequency \citep[$\nu_R=2\nu_0^2/3\nu_B$][]{Ramaty1967,Ramaty1969,Ramaty1968} is then $\sim$ 0.96~GHz.

The broken-power-law distribution function of energetic electrons is expressed as
  \begin{equation}
  N(E)=A E^{\delta_1}(1+(\frac{E_B}{E})^{\delta_1-\delta_2}),
  \label{eq:1}
  \end{equation}
{where $E_B$ represents the break energy (in unit of MeV)}, $\delta_1$ ($\delta_2$) is the spectral index below (above) the break energy. The corresponding single power-law spectrum is represented as
  \begin{equation}
 N(E)=B E^{\delta_1},
  \label{eq:2}
  \end{equation}
where normalization factor B, is related to A in Eq.~\ref{eq:1} by $B=A(1+({E_B}/{0.03})^{\delta_1-\delta_2})$. This gives the same value of electron density at the low energy cutoff (30~keV) for both broken and single power-law spectra. To derive the value of the constant $A$ from the above equation, we set the electron density (per MeV) at 30~keV to be $2\times10^9~$cm$^{-3}$MeV$^{-1}$ {(valid for all cases in this paper)}. The high energy cutoff is taken to be 30~MeV. {For example, the total number density ($N_{nth}$) is 2.68$\times10^7$~cm$^{-3}$ for energetic electrons with a single power-law distribution ($\delta$=-3.0). For electrons with a broken-power-law distribution ($\delta_1=-3.0, \delta_2>\delta_1$), $N_{nth}$ gets larger accordingly (by several percentages to several times)}.{ According to the typical $\delta$ value of -3 to -5 \citep[in some cases larger than -2,][]{Effen2017,Oka2018} and $E_B$ value around 1.0~MeV \citep{Kong2013,White2011}, in our parameter study we vary the three parameters in the following ranges: $-5.0\leq\delta_1\leq-2.0$, 100~keV$\leq{E_B}\leq5.0$~MeV, and $-3.0\leq\delta_2\leq0$.}

Both the size and depth of the source are fixed to be 14$^{\prime\prime}$ ($\sim10^9$~cm). Two viewing angles have been considered, with $\theta=20^\circ$ for the quasi-parallel case and $\theta=70^\circ$ for the quasi-perpendicular case.

With the above parameters, we calculate the emissivity ($\eta_\nu$) and absorption coefficient ($\kappa_\nu$) as a function of frequency (1-100 GHz), for both X and O modes. Note that the transition energy from gyrosynchrotron to synchrotron emission is taken to be 5.11~MeV. The flux density ($I_\nu$) can then be expressed by the solution of radiative transfer equation,
  \begin{equation}
  I_\nu=\int_\Omega S_\nu (1-\exp(-\tau_\nu))d\Omega,
  \label{eq:3}
  \end{equation}
where $S_\nu (= {\eta_\nu}/{\kappa_\nu})$ is the source function, $\tau_\nu (= \kappa_\nu L)$ is the optical thickness, and $\Omega$ and $L$ represent the solid angle and {the geometrical thickness of the source along the line of sight (LOS)}.

\section{NUMERICAL RESULTS}
\label{sec:3}
 In this section, we first compare the results for electrons with a broken-power-law spectrum (see Eq.\ref{eq:1}) to the corresponding single power-law spectrum (see Eq.\ref{eq:2}), {and then study the effect of electron energy on the microwave flux density spectrum.} Finally, we conduct a parameter study to investigate the effect of electron spectra on the microwave emission.

\subsection{Comparison of microwave emission from electrons with power-law and broken-power-law spectra}
\label{sec:3.1}

\begin{figure}
\centering
\epsscale{.9}
\plotone{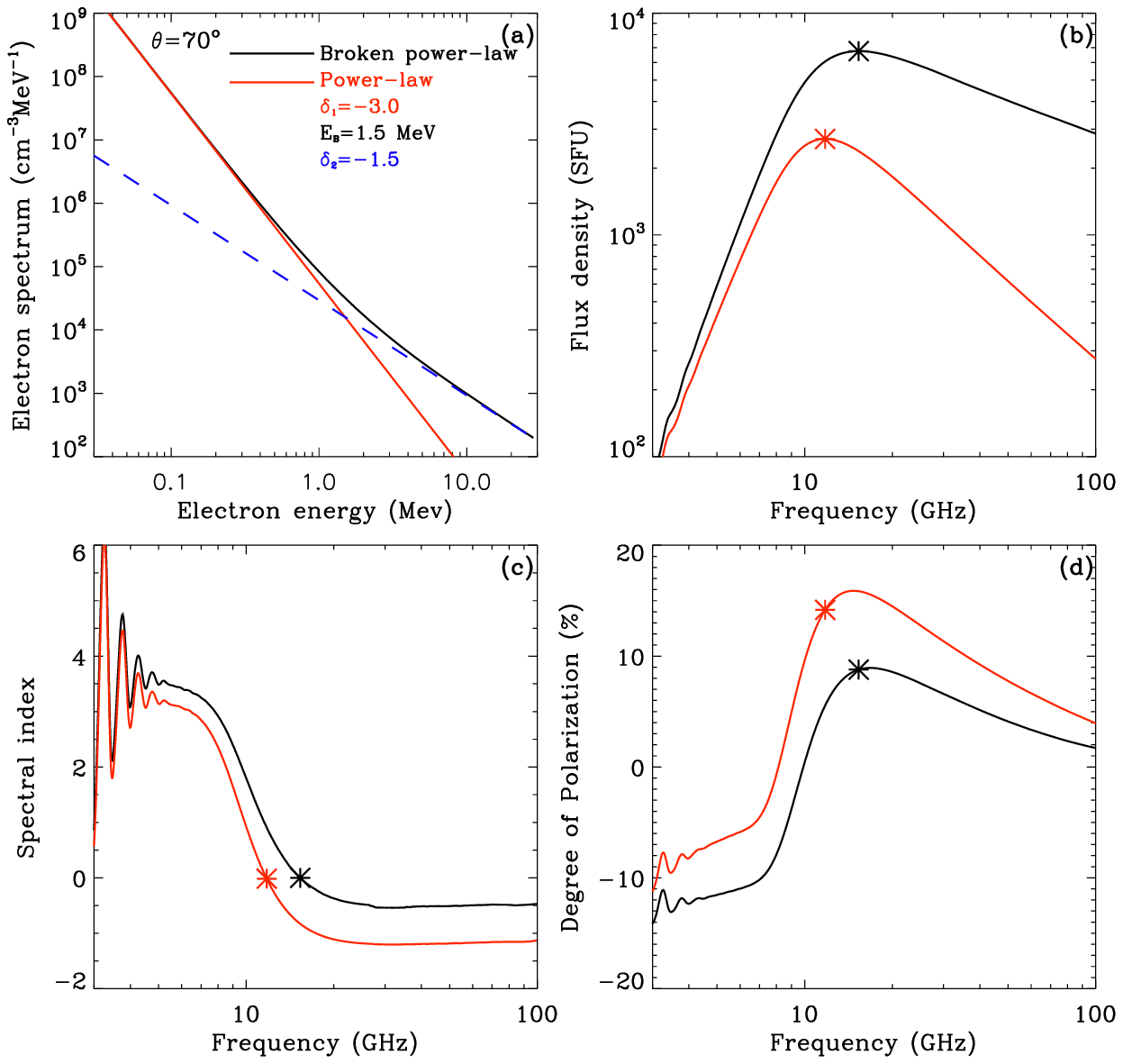}
\caption{Comparison of gyrosynchrotron emission generated by electrons with a single power-law (PL, red) and a broken-power-law (BPL, black) distrbution. Panel (a) shows the electron energy spectra. The broken-power-law spectrum (see Eq.\ref{eq:1}) is shown as the black solid curve, with low energy spectral index $\delta_1=-3.0$ (red solid), break energy $E_B=1.5$~MeV, and high energy spectral index $\delta_2=-1.5$ (blue dashed). For comparison, the red solid curve represents a single power-law spectrum (see Eq.\ref{eq:2}). Panels (b)-(d) present the dependence of the emission flux density, spectral indices ($\alpha$), and the degree of polarization on frequencies (3-100~GHz). The black/red asterisks denotes the turn-over frequency of the corresponding spectrum.}
 \label{fig1-1}
\end{figure}

In Figure \ref{fig1-1}(a), we plot the broken-power-law distribution (black solid, referred to as BPL hereinafter) and the corresponding power-law distribution (red solid, referred to as PL). The blue dashed line shows the component of the high-energy band distribution ($N(E)=AE_B^{\delta_1-\delta_2}E^{\delta_2}$). The parameters are set to be: $\delta_1$=-3.0, $\delta_2$=-1.5, $E_B$=1.5~MeV, and $\theta = 70^\circ$. The calculated microwave flux density ($I_{\nu}$), spectral index ($\alpha_{\nu}$), and degree of circular polarization ($r_c$) for electrons with BPL (black) and PL (red) spectra, in a frequency range of 3-100 GHz, are presented in Figures \ref{fig1-1}(b)-(d). The asterisks denote the peak frequency ($\nu_p$) of the emission (black for BPL, red for PL), which separate the optically-thick and optically-thin regimes.

We see that all radiative quantities are considerably affected by the change of the electron energy distribution. In particular, the total flux density gets significantly larger over all frequencies if comparing the result for BPL to that for PL, and the relative difference increases with frequencies. The peak frequency (${\nu_p}$) increases from about 11.7 GHz for PL to about 15.4 GHz for BPL, and the peak flux density ($I_{\nu_p}$) increases from 2710 SFU to 6743 SFU. In addition, across the whole range of frequencies the spectra are relatively harder for the BPL case than that for the PL case (Figures \ref{fig1-1}(b) and \ref{fig1-1}(c)). In the optically-thick regime, the spectral indices ($\alpha$) are positive and around 3 for PL and 3.5 for BPL (below about 8 GHz), and decrease to negative values of about -1.2 for PL and about -0.5 for BPL. The curves of degree of polarization ($r_c$), as plotted in Figure \ref{fig1-1}(d), move towards more negative values in general (from PL to BPL). {In the optically-thick regime (below $\sim$8~GHz), the range of $r_c$ is about -5\% to -10{\%} for PL and about -10\% to -15{\%} for BPL, while in the optically-thin regime, $r_c$ ranges from about 5\% to 15{\%} for PL and from about 5\% to 10{\%} for BPL}. For both cases, the degree of polarization reaches the maximum near the peak frequency ($\nu_p$).

\begin{figure}
\centering
\epsscale{.9}
\plotone{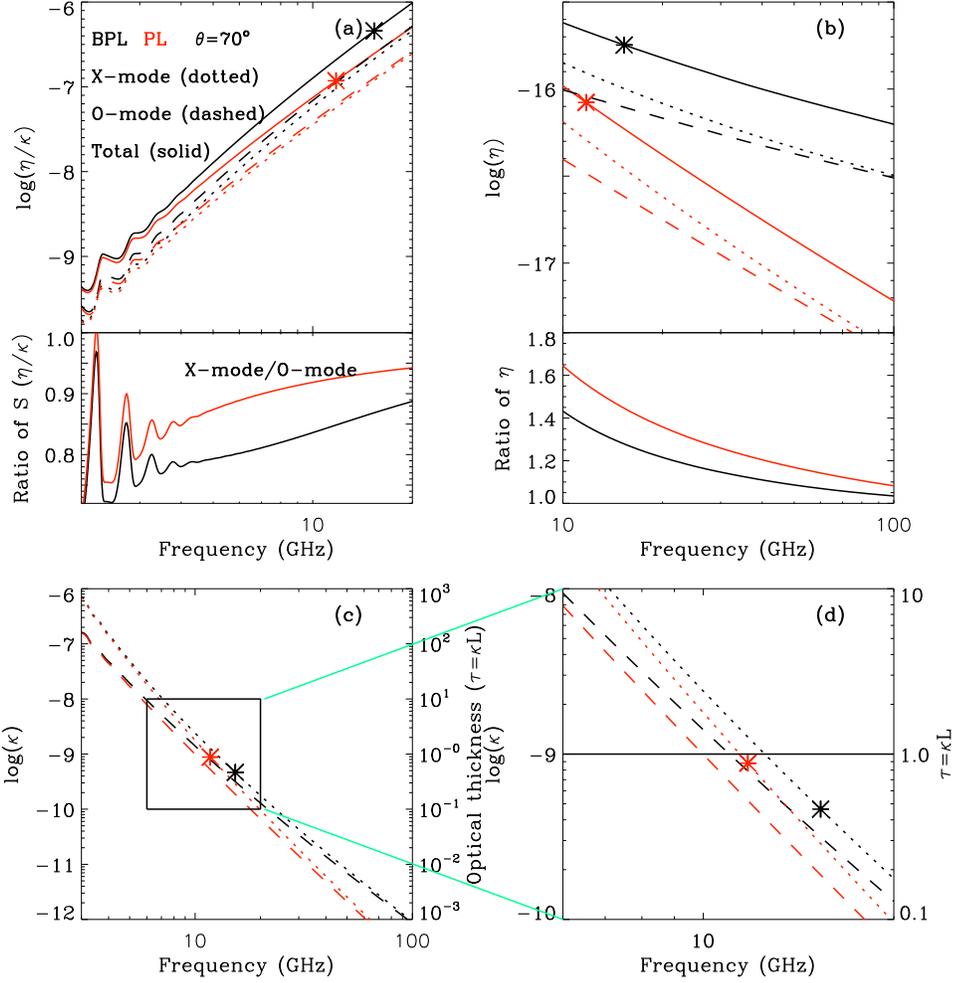}
\caption{
Comparison of the source function $S_\nu$ (a), emissivity $\eta_\nu$ (b), and absorption coefficient $\kappa_\nu$ / optical thickness $\tau_\nu$ (c and d) for X-mode (dotted), O-mode (dashed) and summation of both modes (solid) for BPL (black) and PL (red). In the bottom panels of (a) and (b), black (red) curves show the ratios of parameters between X-mode and O-mode for BPL (PL). The box in the top panel of (c) is enlarged and shown in panel~(d). The black (BPL) and red (PL) asterisks represent the peak frequency derived from the flux density curves (see Figure~\ref{fig1-1}(b)). The parameters of energy spectrum of electrons are the same as those used in Figure~\ref{fig1-1}.}
\label{fig1-3}
\end{figure}

To understand the above results, we plot in Figure~\ref{fig1-3} the source function ($S_\nu=\eta_\nu/\kappa_\nu$), the emissivity ($\eta_\nu$), {and the absorption coefficient ($\kappa_\nu$, optical thickness $\tau=\kappa L$) for X-mode (dotted), O-mode (dashed) and summation of both modes (solid) in the two cases (BPL and PL)}. It is well known that the radiative transfer equation (Eq.\ref{eq:3}) can be simplified as $I_\nu=S_\nu\Omega$ for the optically-thick limit ($\tau \gg 1$) and $I_\nu=\eta_\nu\Omega{L}$ for the optically-thin limit ($\tau \ll 1$), where $I_\nu$ is the total flux density summed over the flux density for both X ($I_{\nu}^X$) and O ($I_{\nu}^O$) modes. Namely, in the optically-thick regime the total flux density is mainly determined by the source function while it is mainly determined by the emissivity in the optically-thin regime. Therefore, the increase of the flux density from PL to BPL can be understood from the plots of $S_\nu$ (Figure 2a) and $\eta_\nu$ (Figure 2b). It can be seen that $S_\nu$ increases by a few times in the optically-thick regime ($\nu < \nu_p$), for instance, at 8 GHz, $\log{S}_\nu= -7.73$ (X-mode), -7.68 (O-mode) {and -7.41 (total)} for PL and $\log S_\nu= -7.55$ (X-mode), -7.46 (O-mode) {and -7.20 (total)} for BPL, while $\eta_\nu$ increases much more (by several times to a few orders in magnitude) in the optically-thin regime, for instance, at 30 GHz, $\log\eta_\nu= -16.9$ (X-mode), -17.0 (O-mode) {and -16.6 (total)} for PL and $\log\eta_\nu= -16.2 $ (X-mode), -16.3 (O-mode) {-15.9 (total)} for BPL. These changes agree with the profiles of the total flux density.

The degree of polarization ($r_c$) is defined as $r_c = ({I_{\nu}^X - I_{\nu}^O})/({I_{\nu}^X + I_{\nu}^O})$. Thus, the overall decline of the $r_c$ curves (see Figure 1d) can also be easily understood from the above plots of $S_\nu$ and $\eta_\nu$ and their ratios between X- and O-modes (see the bottom panels in Figures 2a and 2b). It can be seen that at any specific frequency the relative increases (from PL to BPL) of both $S_\nu$ and $\eta_\nu$ for the O-mode are larger than those for the X-mode. This means the relative enhancement of the O-mode emission across the whole frequency range and thus the general decline of $r_c$. {We also find the differences between X- and O-modes for both $S_\nu$ and $\eta_\nu$ decrease with frequency (see bottom panels in Figure~\ref{fig1-3}(a) and (b)), indicating that $r_c$ changes from negative (positive) values toward zero at low (high) frequencies, which results in the peak of $r_c$ around the turn-over or peak of flux density spectrum.
}

{The microwave spectra usually peak around frequency where optically thickness is approximately 1 due to the increasing $S_\nu$ in the optically-thick limit and decreasing $\eta_\nu$ in the optically-thin limit with frequency (see Figure~\ref{fig1-3}(a) and (b)).} The larger peak frequency resulting from the BPL can be understood by inspecting the plot of $\tau$ (Figure 2c and 2d). It can be seen that for both O- and X-modes the frequencies at $\tau\approx1$ increase from $\sim$10-11 GHz (PL) to 11-13 GHz (BPL), roughly in agreement with the change of $\nu_p$. The increase of the frequency where $\tau\approx1$ is mainly due to the enhanced self-absorption effect as a result of the increase of the density of energetic electrons for the BPL case. Note that the Razin effect is not important due to the rather-low Razin frequency (0.96 GHz).

\subsection{Parameter study}
\label{sec:3.2}
\subsubsection{Emission parameters for electrons with different energy bands}
\label{sec:3.2.0}
\begin{figure}
\centering
\epsscale{.9}
\plotone{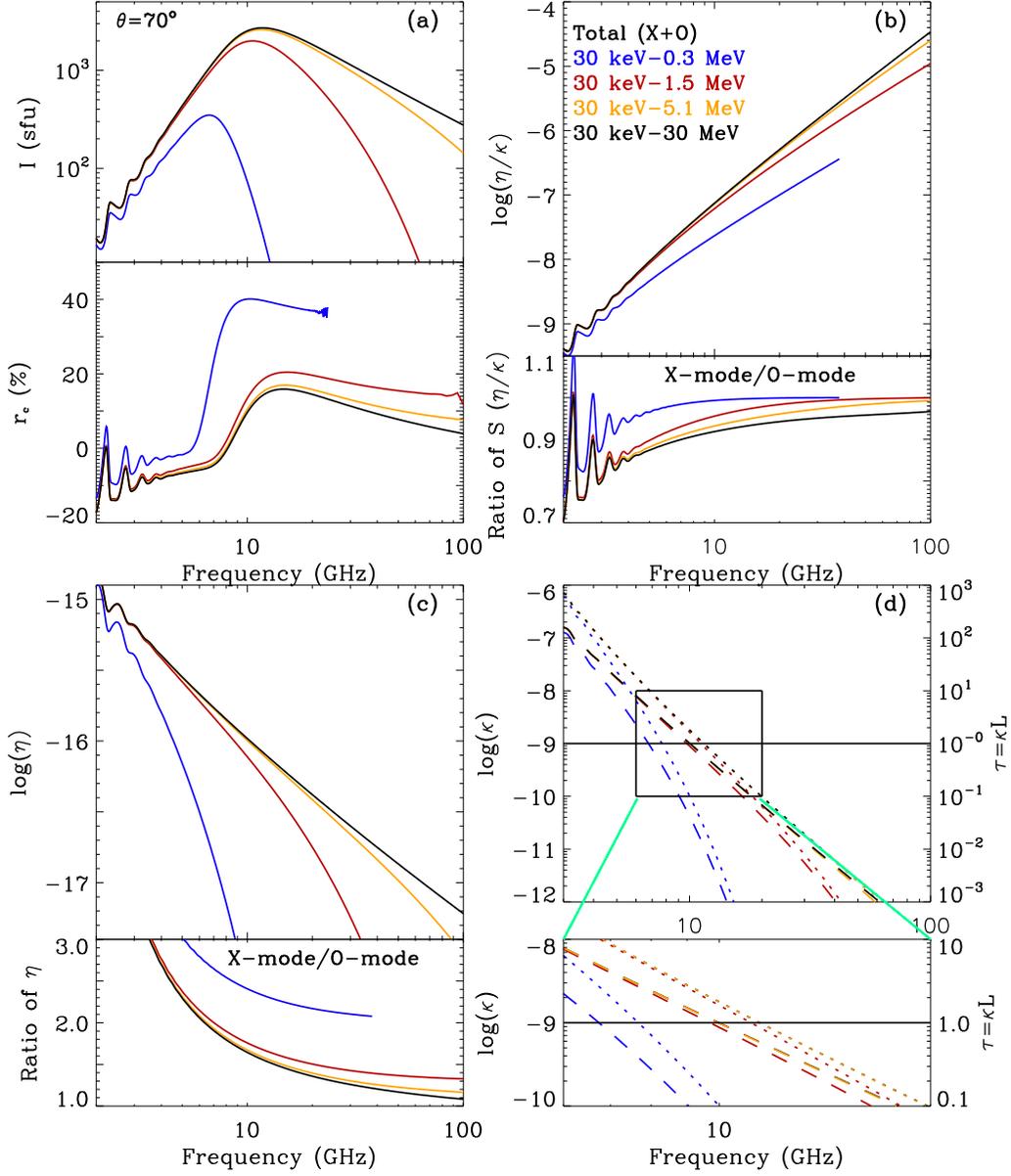}
\caption{
Comparison of the flux density $I_\nu$ \& degree of polarization $r_c$ (a), source function $S_\nu$ (b), emissivity $\eta_\nu$ (c), and absorption coefficient $\kappa_\nu$ / optical thickness $\tau_\nu$ (d) for electrons in different energy ranges (blue for 30~keV-0.3~MeV, red for 30~keV-1.5~MeV, orange for 30~keV-5.1~MeV and black for 30~keV-30~MeV), with solid lines for summation of both modes, dotted lines for X-mode, and dashed lines for O-mode. Bottom panels of (b) and (c) show the ratios of parameters between X-mode and O-mode. The box in the top panel of (d) is enlarged and shown in the bottom panel. The parameters of electron energy distribution are the same as PL used in Figure~\ref{fig1-1}.}
\label{fig1-4}
\end{figure}

{To better understand the above results, we further conduct a study on the effect of electron energy on microwave emission. In Figure \ref{fig1-4}, we plot, as functions of frequency, the dependence of the flux density and the degree of polarization (a), the source function (b), the emissivity (c), and the absorption coefficient / optical thickness (d) calculated with nonthermal electrons (PL, $\delta=-3.0$) in four energy ranges (30~keV-0.3~MeV (referred to as band-I, blue), 30~keV-1.5~MeV (band-II, red), 30~keV-5.1~MeV (band-III, orange), and 30~keV-30~MeV (band-IV, black)).  }

{The flux density changes notably with electron energy ranges (see Figure~\ref{fig1-4}(a)). In the optically-thick regime, $I_\nu$ and $S_\nu$ increase by several percentage to several times with additional electrons ranging from 0.3 to 1.5 MeV (blue to red plots in Figure~\ref{fig1-4}(a) and (b), $N_{nth}^{ 0.3-1.5MeV}\approx3.2\times10^5$~cm$^{-3}$), while they change very slightly with increased number of electrons in the energy range of 1.5-30~MeV (red, orange to black plots). For instance, $I_\nu$ at 6~GHz are $\sim$319, 697, 755 and 766 SFU for electrons in band-I, band-II, band-III, and band-IV, respectively. In the optically-thin regime, $I_\nu$ and $\eta_\nu$ increase significantly with the increasing high-energy cutoff (see Figure~\ref{fig1-4}(a) and (c)). The frequency where $I_\nu$ equals 300~SFU increases from $\sim$7.5, 28.7, 66.5 to 94.0~GHz when the high-energy cutoff changes from 0.3, 1.5, 5.1 to 30~MeV. The above analysis indicates that microwave emission is produced by the electrons below a few MeV at lower frequencies and by electrons with energy above several MeV at higher frequencies. Therefore, the change of flux density from PL to BPL essentially results from the increase of high-energy electron density. As seen from Figure~\ref{fig1-4}(d), the peak frequency is also sensitive to electrons in the range of hundreds keV to a few MeV as they dominate the emissivity and absorption coefficients below $\sim$10-20~GHz}

{Similar to the energy dependence of the flux density, the degree of polarization ($r_c$) is also sensitive to the change of high-energy cutoff. Electrons in the range of 0.3-1.5~MeV reduce the ratio of $S_\nu^X$ to $S_\nu^O$ by the largest extent at low frequencies (see blue and red plots in the bottom panel of Figure~\ref{fig1-4}(b)). At high frequencies (see bottom panel of Figure~\ref{fig1-4}(c)), the ratio of $\eta_\nu$ (between X- and O-modes) decreases with increasing high-energy cutoff. Therefore, at a given frequency the larger cutoff results in a smaller value of the ratio ($\eta_\nu^X $ to $\eta_\nu^O$). This means relatively larger enhancement for O-mode emission is generated by additional high-energy electrons. For instance, ratios of $\eta_\nu^X $ to $\eta_\nu^O$ at 35~GHz is 1.40, 1.27, and 1.23 for electrons in band-II, band-III, and band-IV respectively, which yield $r_c$ to be 16.7\%, 11.9\%, and 10.3\%. }

\subsubsection{Effect of the low energy spectral index $\delta_1$}
\label{sec:3.2.1}
To study the influence of the low energy spectral index $\delta_1$, we vary $\delta_1$ from -5.0 to -2.0 with other parameters fixed ($E_B=1.5$ MeV, and $\delta_2= -1.5$) {for nonthermal electrons ranging from 30~keV-1.5~MeV (dashed) and 30~keV-30~MeV (solid)}. The resulting $I_\nu$, $\alpha$, and $r_c$ for three values of $\delta_1$ (-4.0 (red), -3.5 (blue), and -3.0 (black)) are shown in Figure \ref{fig2} for the quasi-parallel ($\theta = 20^\circ$, left panels) and quasi-perpendicular ($\theta = 70^\circ$, right panels) viewing angles. We can see that for both viewing angles the increase of $\delta_1$ results in enhanced $I_\nu$ (solid lines). In the optically-thick regime $I_\nu$ increases by a factor of 2-10, and in the optically-thin regime $I_\nu$ increases by up to 2 orders of magnitude, when $\delta_1$ changes from -4.0 to -3.0. As seen from the upper and middle panels of this figure, the spectral index $\alpha$ almost keeps constant in the high frequency part ($>$15 GHz for $\theta=20^\circ$ and $>$30 GHz for $\theta=70^\circ$), while $\alpha$ varies significantly in the corresponding low frequency part of the spectrum. For instance, the spectra at low energy present a soft-hard variation (see the red curve) in the frequency range of [3, 15] GHz for $\theta=20^\circ$ and [5, 30] GHz for $\theta=70^\circ$.

As seen from the upper panels, the peak frequency ($\nu_p$) also increases with increasing $\delta_1$, being $\sim$4.0, 6.1, and 10.6~GHz for $\theta=20^\circ$ and 6.8, 9.6, and 15.4~GHz for $\theta=70^\circ$, for the above values of $\delta_1$.

Figures $\ref{fig2}$(e) and (f) show that similarly to the $\delta_1$ dependence of the spectral index, the degree of polarization $r_c$ (solid) is also sensitive to $\delta_1$ in the low frequency part (below $\sim$15~GHz for $\theta=20^\circ$ and below $\sim$30~GHz for $\theta=70^\circ$) but remains unchanged in the high frequency part. With increasing $\delta_1$, $r_c$ decreases in general. As mentioned in the above subsection, this is due to the relative enhancement of the O mode emission (in comparison with that of the X mode). Again, the maxima of $r_c$ always appear around $\nu_p$ and decrease from about 85\% ($\delta_1=-4.0$) to 45\% ($\delta_1=-3.0$) for $\theta=20^\circ$, and from about 23\% to 9\% for $\theta=70^\circ$.

{To better understand the above changes, we compare the emission from electrons in the range of 30~keV-1.5~MeV (dashed) with that from electrons of 30~keV-30~MeV (solid) in Figure~\ref{fig2}. It is clear that the enhanced flux density, increased spectral index and decreased degree of polarization beyond several GHz are indeed produced by the additional electrons ranging from 1.5~MeV to 30~MeV. Low-frequency emission is mostly produced by the electrons below 1.5~MeV, which almost has no contribution to high-frequency emission and thus the polarization as shown in the bottom panels.}

\subsubsection{Effect of the break energy $E_B$}
\label{sec:3.2.2}
One may expect that if $E_B$ increases (with other parameters fixed), the effect of the high-energy component of the BPL shall become less significant. This is consistent with our results shown in Figure~\ref{fig3}, with $E_B$ taken to be 0.5 (black), 1.5 (blue), and 2.5 (red) MeV, respectively, and $\delta_1=-3.0$, $\delta_2=-1.5$. {Note that for comparison, emission parameters produced by electrons with energies below each $E_B$ are plotted as dashed lines. }

It can be seen that for the two viewing angles, the variations of the three sets of curves are similar to each other. Both the flux density $I_\nu$ and the peak frequency $\nu_p$ decrease with increasing $E_B$. The variation of $I_\nu$ in the optically-thick regime is less pronounced than that in the optically-thin regime. For instance, for both viewing angles $I_\nu$ increases by more than 1 order of magnitude as $E_B$ falls from 2.5~MeV (red) to 0.5~MeV (black) in the optically-thin regime, but in the optically-thick regime the relative increases of $I_\nu$ are in general less than 5-6 times. Note that at frequencies above $\nu_p$ the spectral indices only weakly depend on $E_B$ (see also the middle panels of Figure $\ref{fig3}$). $\nu_p$ is close to or smaller than 10~GHz when $E_B$ is larger than $\sim$1.5~MeV ($\theta=20^\circ$) or $\sim$2~MeV ($\theta=70^\circ$).

The degree of polarization ($r_c$), as presented in the lower panels of Figure $\ref{fig3}$, increases with increasing $E_B$. The maxima of $r_c$ are $\sim$44 (8.3), 29 (5.4), and  10 (2.7)\% for $E_B$=2.5, 1.5, and 0.5~MeV, respectively, for the quasi-parallel (perpendicular) case. At higher frequencies the dependence of $r_c$ on $E_B$ is weaker.

{
Emission parameters for electrons below break energy (dashed lines in Figure~\ref{fig3}) and high-energy cutoff (solid lines) are further compared. It is clear that electrons in the range of break energy to high-energy cutoff determine microwave emission at higher frequencies (typically above 10~GHz). In addition, we can see that there is obvious dependence of emission parameters on electron energy bands. For instance, the optically-thin flux density decreases to 100 (500) SFU at about 5.7 (10.9), 14.1 (29.0), and 20.6 (41.7) GHz for electrons with energy range of 0.03-0.5~MeV, 0.03-1.5~MeV, and 0.03-2.5~MeV, respectively, at viewing angle of $\theta=20^{\circ}$ (70$^{\circ}$).
}

\subsubsection{Effect of the high energy spectral index $\delta_2$}
\label{sec:3.2.3}
In Figure~\ref{fig4}, we plot the results obtained by varying $\delta_2$ to be -1.0 (black), -1.5 (blue), and -2.0 (red), and fixing $\delta_1=-3.0$, $E_B=1.5$~MeV. Again, the overall dependence of the three parameters (solid lines) on $\delta_2$ is similar for the two viewing angles. The upper panels show that $I_\nu$ depends sensitively on $\delta_2$ in the optically-thin regime, while the dependence is much weaker in the optically-thick regime. Both the peak frequency $\nu_p$ and the peak flux density $I_p$ increase with increasing $\delta_2$. For instance, $\nu_p$ and $I_p$ are 8.6 (14.2) GHz and 1193 (5157) SFU for $\delta_2 = -2.0$, and become 15.0 (17.6) GHz and 3155 (10887) SFU for $\delta_2 = -1.0$, respectively, in the viewing angle of $\theta=20^\circ~(70^\circ)$.

Top and middle panels of Figure~\ref{fig4} show that the spectral index $\alpha$ does not change significantly in the optically-thick regime, but increases (i.e., the spectra get harder) with increasing $\delta_2$ in the optically-thin regime. For instance,  $\alpha$ is in the range of [-0.75, -0.70] for $\delta_2 = -2$ and  [-0.35 to -0.25] for $\delta_2 = -1$ (for both viewing angles). This dependence of $\alpha$ on $\delta_2$ is different from that on the other two parameters ($\delta_1$ and $E_B$). {For single power-law electron energy distribution, the following relation between the gyrosynchrotron spectral index ($\alpha$) and the spectral index ($\delta$) of energetic electrons has been wildly used \citep{Dulk1982,Dulk1985}}
\begin{equation}
\alpha=1.22+0.9\delta,
\label{eq:4}
\end{equation}
{This is consistent with our results that for broken power-law distribution} $\alpha$ in the high-frequency, optically-thin regime is mainly determined by the spectral index of electrons at high energy ($\delta_2$). {We note that the above relation is derived under certain range of parameters (see the introduction) and deviates somewhat from strict calculations \citep[see our results in Figure~\ref{fig4}(c) and (d), ][]{Dulk1985,Huang2009,Song2016}.}

The degree of polarization $r_c$ has a sensitive dependence on $\delta_2$ over almost the whole range of frequencies, as seen from Figures \ref{fig4}(e) and (f). In general, $r_c$ decreases with increasing $\delta_2$. In the optically-thick regime, $r_c$ increases from negative to positive values and reaches the maxima near the peak frequency $\nu_p$. $r_c$ then decreases in overall in the optically-thin regime. Specifically, at $\theta=20^\circ$ $(70^\circ)$, $r_{cmax}=55\%$ $(12\%)$ for $\delta_2 = -2.0$, and $r_{cmax}=29\%$ $ (5.5\%)$ for $\delta_2 = -1.0$. The frequency dependence of $r_c$ is consistent with the results presented above.

{Again, the parameter comparison for electrons in various energy bands (solid and dashed plots) demonstrates the dominant role of high-energy electrons in high-frequency emission. For instance, $I_\nu$ decreases to below 10~SFU and $r_c$ ranges from 70\% to 85\% (15\% to 20\%) above $\sim$22 (70)~GHz for low-energy electrons (30~keV-1.5~MeV), while $I_\nu$ can still reach up to hundreds to thousands of SFU and $r_c$ is below 35\% (4\%) with the additional electrons above 1.5~MeV, at the viewing angle of $\theta=20^\circ$ $(70^\circ)$.  }

\section{SUMMARY AND DISCUSSION}
\label{sec:4}
  Microwave and HXR emissions emitted by energetic electrons in the solar atmosphere contain valuable information on the background coronal plasmas, magnetic field and nonthermal energetic electrons. The HXR spectrum sometimes presents a hardening feature at energy above 300~keV, which has been interpreted in terms of electron energy distribution in the form of broken-power-law. Starting from this interpretation, we have conducted a parameter study to investigate the effect of broken-power-law spectra of energetic electrons on microwave emissions on the basis of the classical gyrosynchrotron mechanism. The aim of the study is not only to understand how such electron energy distribution affects the microwave emission but to provide insight into some uncommon features of microwave spectra reported lately, such as unusually hard or even positive spectral slope, and/or a super-high peak frequency of microwave spectrum.

  Three parameters of the electron spectrum are of particular interest, including the break energy ($E_B$), the spectral indices below ($\delta_1$) and above ($\delta_2$) the break.
  Comparing the results for the broken to those for the corresponding single power-law spectrum (i.e., with the spectral index $\delta_1$), the following conclusions were reached. (1) The total flux density can increase by several times in the optically-thick regime, and by orders of magnitude in the optically-thin regime; (2) the peak frequency ($\nu_p$) also increases and can reach up to tens of GHz; and (3) the degree of polarization ($r_c$) decreases in general. Parameter study shows that (1) the variation of the flux density is much larger in the optically-thin regime than that in the optically-thick regime, and the microwave spectra around the peak frequency manifest various profiles with the softening or soft-hard pattern; (2) the parameters $\delta_1$ and $E_B$ affect the microwave spectral index ($\alpha$) and the degree of polarization ($r_c$) mainly in the optically-thick regime and around the peak regime, while the effect of $\delta_2$ mainly appears in the optically-thin regime; (3) for different viewing angles, the dependence of the microwave spectral parameters is similar.

\begin{figure}
\centering
\epsscale{.9}
\plotone{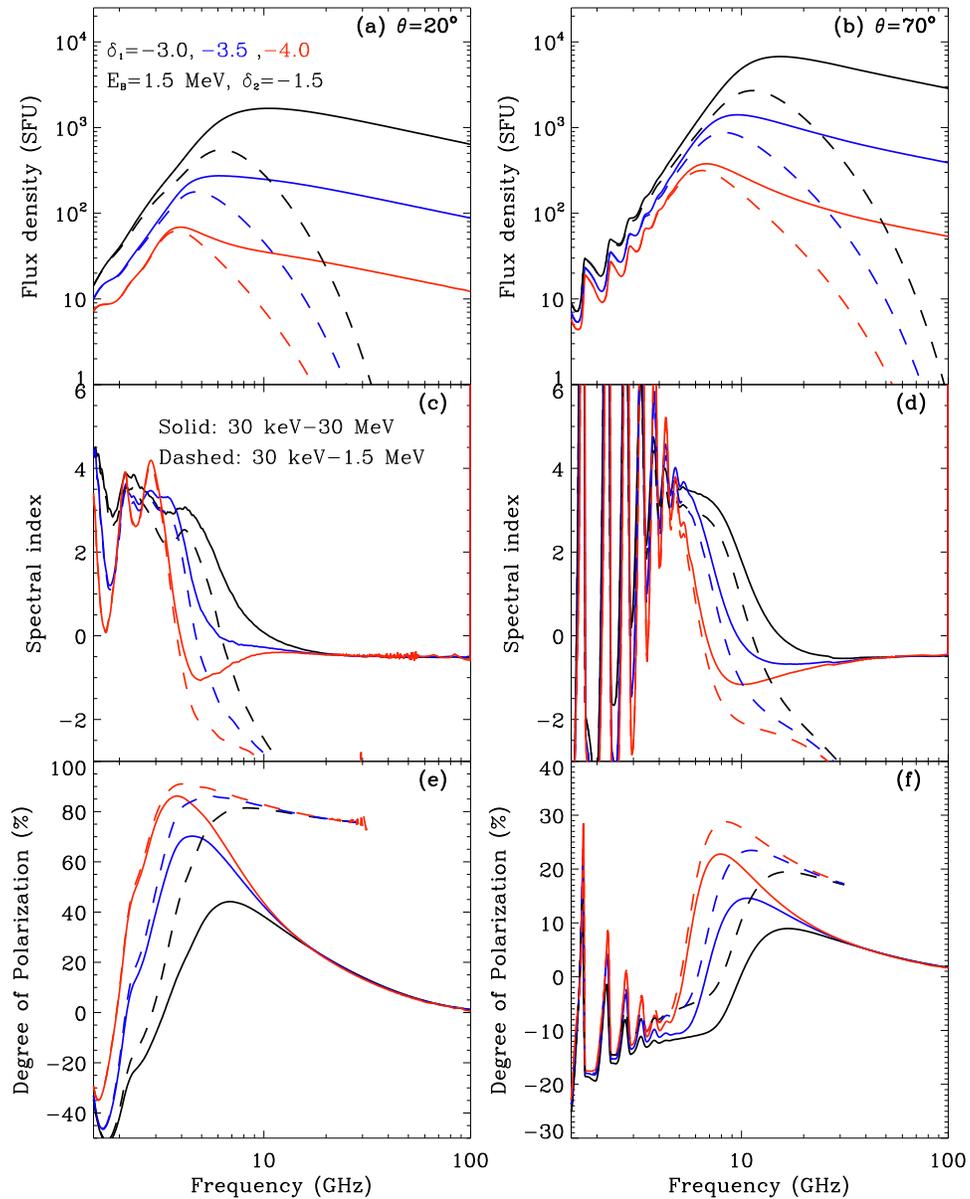}
\caption{Effect of low energy index $\delta_1$ on the microwave emission in the viewing angle of $\theta=20^\circ$ (left) and $\theta=70^\circ$ (right). From top to bottom: total flux (including X-mode and O-mode), spectral indices and degree of polarization vs frequency for different $\delta_1$ ($\delta_1 = -3.0$ (black), -3.5 (blue) and -4.0 (red)), and for $E_B$ = 1.5~MeV, and $\delta_2$ = -1.5. Electron energy ranges: 30~keV-30~MeV (solid) and 30~keV-1.5~MeV (dashed).
}
 \label{fig2}
\end{figure}

\begin{figure}
\centering
\epsscale{.9}
\plotone{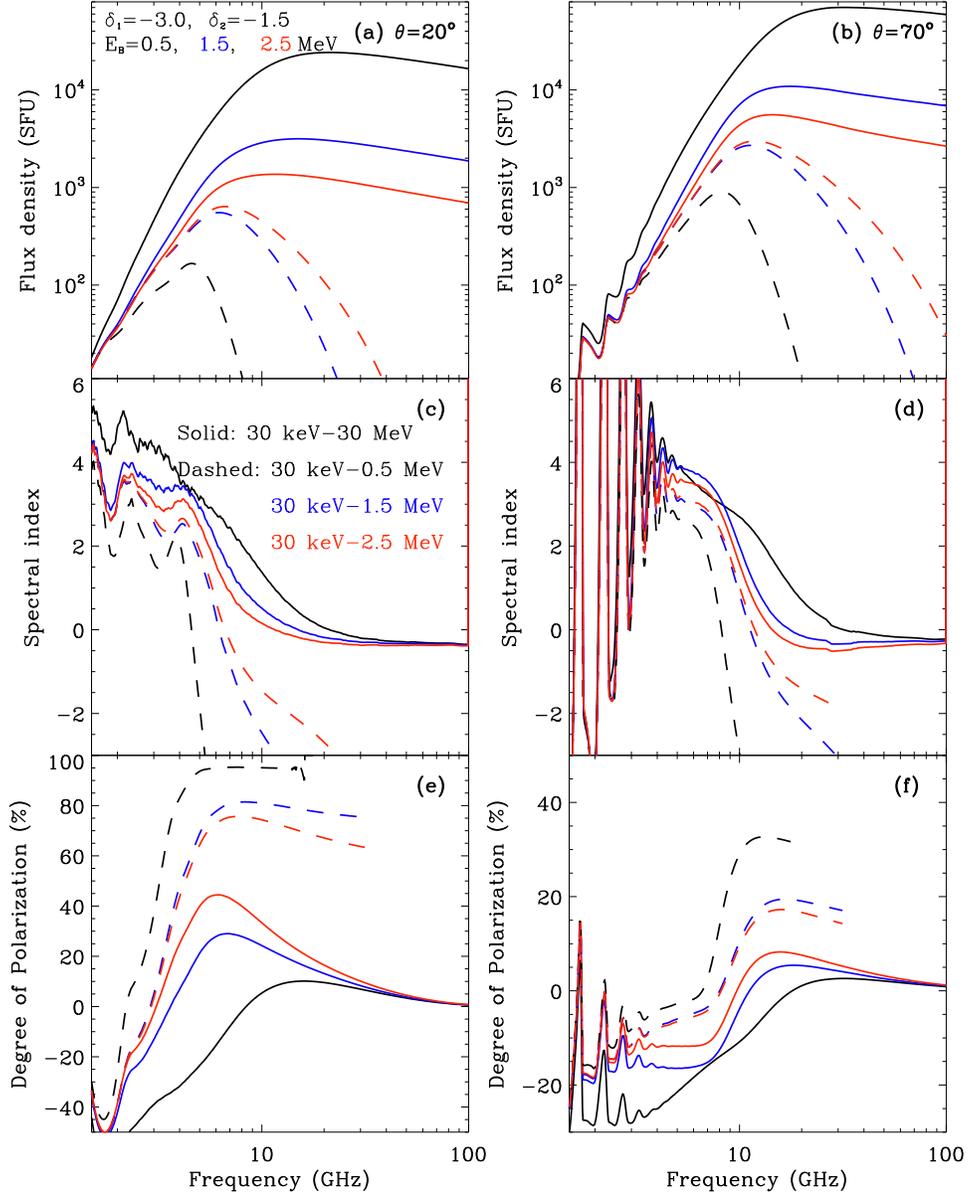}
\caption{Effect of break energy $E_B$ on the microwave emission in the viewing angle of $\theta=20^\circ$ (left) and $\theta=70^\circ$ (right). From top to bottom: total flux (including X-mode and O-mode), spectral indices and degree of polarization vs frequency for different break energy $E_B$ ($E_B=0.5$~MeV (black), 1.5 MeV (blue), and 2.5~MeV (red)), and for $\delta_1=-3.0$, and $\delta_2=-1.5$.  Electron energy ranges: 30~keV-30~MeV (solid), 30~keV-0.5~MeV (black dashed), 30~keV-1.5~MeV (blue dashed) and 30~keV-2.5~MeV (red dashed).
}
 \label{fig3}
\end{figure}

\begin{figure}
\centering
\epsscale{.9}
\plotone{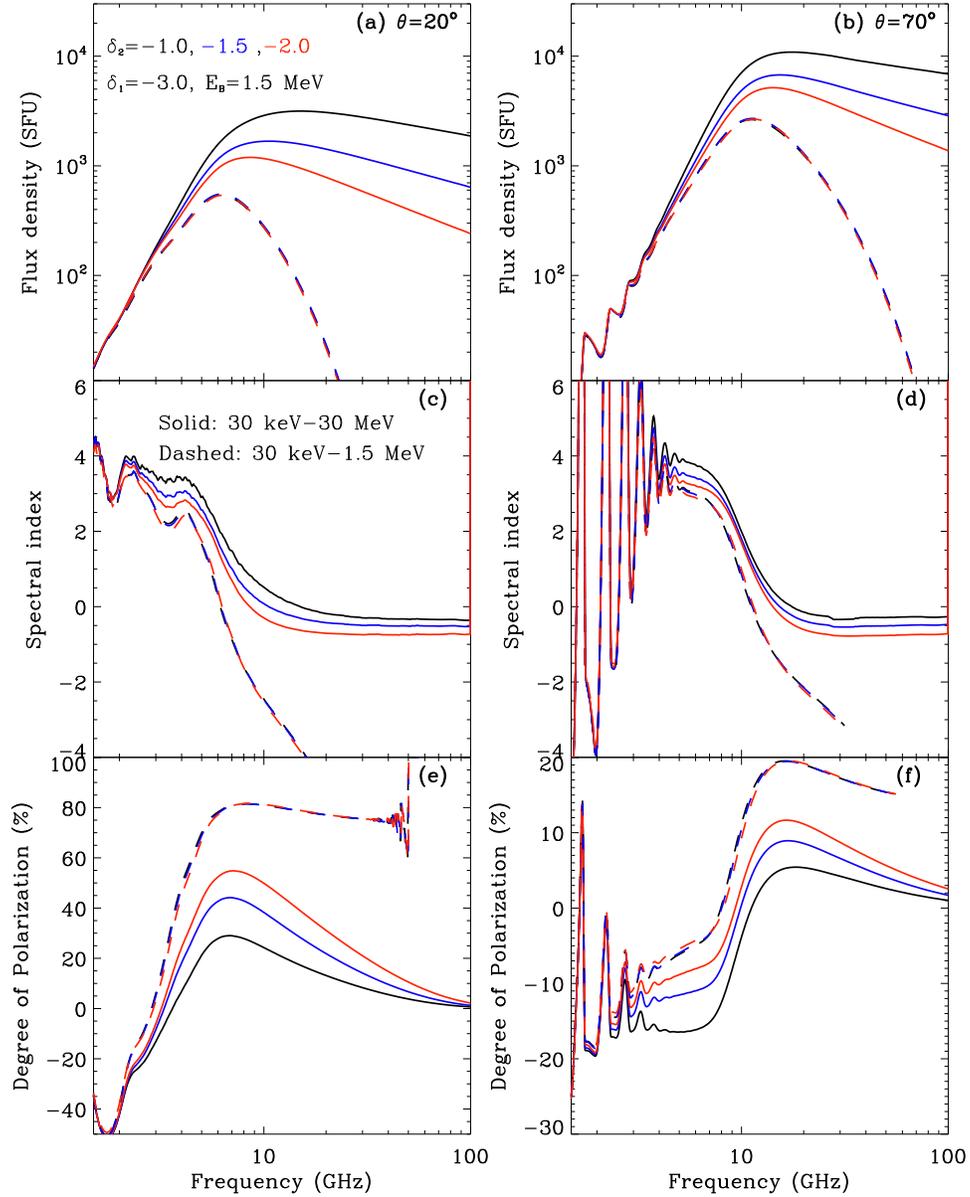}
\caption{Effect of high energy index $\delta_2$ on the microwave emission in the viewing angle of $\theta=20^\circ$ (left) and $\theta=70^\circ$ (right). From top to bottom: total flux (including X-mode and O-mode), spectral indices and degree of polarization vs frequency for different high energy spectral indices $\delta_2$ ($\delta_2$=-1.0 (black), -1.5 (blue), and -2.0 (red)), and for $\delta_1=-3.0$, and $E_B=1.5$~MeV. Electron energy ranges: 30~keV-30~MeV (solid) and 30~keV-1.5~MeV (dashed).
}
 \label{fig4}
\end{figure}

  Some of our results can be understood in a qualitative way. According to our calculations, the microwave emission in the low-frequency part is mainly contributed by the relatively low-energy electrons and vice versa. This explains why the effects of $\delta_1$ and $E_B$ on the microwave emission mainly appear in the low-frequency part while the effect of $\delta_2$ mainly takes place in the high-frequency part. The change of $r_c$ with electron energy distribution is mainly due to the relative intensity of the O-mode emission and the X-mode emission. According to \citet{Ramaty1969} and \citet{Fleishman2003a}, the O-mode emission (left polarized in this present configuration) is generated by energetic electrons with energy higher than that for the X-mode (right polarized). This explains why we find $r_c$ declines in general with the hardening of the electron spectra. We also notice that $r_c$ for the quasi-parallel viewing angle is larger than for the quasi-perpendicular viewing angle (Figures~\ref{fig2}-\ref{fig4}). This is due to the fact that the O-mode emission presents a radiative cone (centered around the perpendicular direction of magnetic field lines) narrower than that of the X-mode emission \citep{Ramaty1969,Fleishman2003a}. Thus, the larger the viewing angle, the stronger the O-mode emission one can observe.

  Earlier observational reports of solar flares reveal some uncommon features of microwave spectra, such as unusually hard (or even positive) spectra, and/or a super-high peak frequency reaching up to tens of~GHz \citep[e.g.][]{Hachenberg1961,Klein1987}. {In relevant studies \citep[e.g.][]{Ramaty1969, Klein1987,Raulin1999,Kaufmann2004,Luthi2004,Silva2007,Trottet2011,Wu2016,Song2016}, these observations have been explained in terms of enhanced Razin suppression (with extremely high density of ambient plasmas), enhanced self-absorption effect (assuming a high density of energetic electrons), or enhanced absorption by cool plasmas outside the source. For example, it has been proposed that $\nu_p$ of tens of GHz is due to the Razin effect (self-absorption) when the thermal (nonthermal) electron density is over 10$^{11}$~cm$^{-3}$ (10$^9$~cm$^{-3}$) for magnetic filed of several hundred gauss \citep{Klein1987,Klein2010,Melnikov2008,Fleishman2011,Fleishman2017,Grechnev2017}. However, in most of these studies a single power-law energy distribution of non-thermal electrons was assumed. Our study based on broken-power-law energy distribution shows that the increased high-energy electrons specified by higher $\delta_2$ can result in enhanced flux density spectrum that hardens at higher frequencies, even though the total number density of nonthermal electrons does not change much around $10^7$~cm$^{-3}$. }The obtained microwave spectra do manifest, in many cases, the peak frequencies as high as a few tens of GHz, and spectral indices as flat as -0.25. For instance, $\nu_p=16.6$ (21.3) GHz and $\alpha=-0.31$~(-0.25) for $\delta_1=-3.0$, $E_B$=1.0~MeV, $\delta_2=-1.0$, $N_{nth}=2.73\times10^7$~cm$^{-3}$, and $\theta=20^\circ$ $(70^\circ)$. Thus, our study provides a novel line of thought in understanding relevant observations.

  As shown by our results \citep[and those published earlier, e.g., ][]{Ramaty1969,Klein1987,Fleishman2003a,Fleishman2003b,Melnikov2008}, all relevant parameters (such as the microwave intensity, the spectral index, the degree of polarization) vary significantly with frequency, especially around the peak frequency. The peak frequency varies from a few to tens of GHz. This poses a serious challenge to present solar radio spectrographs and heliographs. Most of current instruments work at frequencies below 18 GHz such as the newly-updated Expanded Owens Valley Solar Array \citep[EOVAS,][]{Gary2018}, or at few discrete frequencies such as Nobeyama Radio Polarimeters \citep[NoRP,][]{Nakajima1985} and Nobeyama RadioHeliograph \citep[NoRH,][]{Nakajima94}. Thus, the data available cannot fully account for the continuous variation of the microwave spectra with frequencies. Construnction of next generation radio telescopes that can provide a more-complete spectral coverage of microwave bursts, especially, in the high-frequency part, say, $>10-30$ GHz is desirable.

\acknowledgments
This work is supported by the National Natural Science
Foundation of China 11703017, 41331068, 11790304 (11790300), 11503014, 11873036,
and Shandong Provincial Natural Science Foundation, China ZR2016AP13. X.K. also acknowledges the Young Scholars Program of Shandong University, Weihai.

\end{document}